\documentclass[%
 reprint,
superscriptaddress,
nolongbibliography,
 amsmath,amssymb,
 aps,
pra,
]{revtex4-2}

\usepackage{graphicx}
\usepackage{dcolumn}
\usepackage{bm}
\usepackage{hyperref}
\usepackage[mathlines]{lineno}
\usepackage{epstopdf}
\usepackage{color}
\usepackage{dsfont}
\usepackage{mathtools}
\usepackage{lipsum}
\usepackage{cuted}
\usepackage{amssymb}
\usepackage{hyperref}
\usepackage{url}

\def\Xint#1{\mathchoice
{\XXint\displaystyle\textstyle{#1}}%
{\XXint\textstyle\scriptstyle{#1}}%
{\XXint\scriptstyle\scriptscriptstyle{#1}}%
{\XXint\scriptscriptstyle\scriptscriptstyle{#1}}%
\!\int}
\def\XXint#1#2#3{{\setbox0=\hbox{$#1{#2#3}{\int}$}
\vcenter{\hbox{$#2#3$}}\kern-.5\wd0}}

\def\dashint{\Xint-}

\begin{document}




\title{Exploring the equivalence of causality-based and quantum mechanics-based sum rules for harmonic generation in nonlinear optical materials}



\author{Theodoros T. Koutserimpas}
 \affiliation{School of Electrical and Computer Engineering, Cornell University, Ithaca, New York 14853, USA}
 
 \author{Hao Li}
 \affiliation{Department of Applied Physics and Energy Sciences Institute, Yale University, New Haven, Connecticut 06511, USA}

 \author{Owen D. Miller}
 \affiliation{Department of Applied Physics and Energy Sciences Institute, Yale University, New Haven, Connecticut 06511, USA}

\author{Francesco Monticone}

 \affiliation{School of Electrical and Computer Engineering, Cornell University, Ithaca, New York 14853, USA}

\begin{abstract}
The Kramers-Kronig relations and various oscillator strength sum rules represent strong constraints on the physical response of materials. In this work, taking inspiration from the well-established equivalence between $f-$sum rules and Thomas--Reiche--Kuhn sum rules in linear optics, we explore the connection between causality-based and quantum-mechanics-based sum rules in the context of nonlinear optical processes. Specifically, by considering the sum-over-states expression for the second harmonic generation susceptibility, we deduce a new representation basis for the imaginary part of this susceptibility and we use it to derive, from causality-based integral sum rules, a new set of discrete sum rules that the transition dipole moments must satisfy. As in the case of the Thomas--Reiche--Kuhn sum rules, we also show that these results can alternatively be derived through an independent quantum mechanical analysis. Finally, we consider the implications of the derived sum rules for the second-harmonic-generation susceptibility of two- and three-level systems and, more broadly, we discuss the possible significance and challenges of using these results for the goal of identifying fundamental limits to the response of nonlinear optical materials.
\end{abstract}

\maketitle


\section{\label{sec:intro}Introduction}

Any physical response must satisfy the principle of causality \cite{Toll}. In linear optics, this principle dictates that the electric susceptibility of a material, $\chi^{(1)}(\omega)$, is holomorphic in the upper half of the complex $\omega-$plane, where $\omega$ is the angular frequency of the electromagnetic excitation. As a result of this property, and assuming the susceptibility is sufficiently well-behaved (square integrable), ${\mathop{\rm Re}\nolimits} \{ {{\chi ^{(1)}}(\omega )} \}$ and ${\mathop{\rm Im}\nolimits} \{ {{\chi ^{(1)}}(\omega )} \}$ are connected through integral relations known as the Kramers--Kronig relation \cite{Nussenzveig}. However, causality does not always lead to analyticity in nonlinear optics, as seen for example in the nonlinear process of self-induced change in refractive index \cite{Boyd2020}. 

Nevertheless, Kramers--Kronig relations can be established for many nonlinear optical responses of  interest, including any process in which all input frequencies are fixed (corresponding to fixed pump beams, e.g.), except one (the "probe" beam) \cite{Lucarini}. For cases in which the input frequencies are mutually dependent and varied at the same time, the Bassani--Scandolo theorem \cite{Bassani1991} gives  conditions under which Kramers-Kronig relations can  be derived. This theorem relies on the selection of a one-dimensional space embedded in the $n$-dimensional space of the interdependent frequency variables. Notably, harmonic generation, of any order, satisfies the Bassani--Scandolo theorem and Kramers--Kronig relations can be derived from the holomorphic properties of the corresponding susceptibility \cite{Lucarini}.

A prominent nonlinear process satisfying Kramers--Kronig relations is second-harmonic generation (SHG), described by a second-order susceptibility ${\chi ^{(2)}}(\omega ,\omega )$, which results in the doubling of the incident frequency of light. Dipoles of noncentrosymmetric materials can oscillate under the action of the incident frequency $\omega$ as asymmetric anharmonic oscillators, radiating at $2\omega$, twice the incident frequency $\omega$, in addition to their usual radiation at $\omega$. This nonlinear process is employed in high-power lasers at specific wavelengths, as well as  various applications in microscopy, ultra-short pulse measurement, and materials characterization \cite{Boyd1965,Kumar2009,Moreaux2000,Sidick1995,Sidick1995a,Ren2015}. Despite being one of the most important processes in nonlinear optics, there is still limited understanding of the fundamental constraints to harmonic generation in terms of either its resonance properties (e.g., optimal number and  location of resonances) or the involved transition dipole moments. The most important known constraint on transition dipole moments is the so-called Thomas--Reiche--Kuhn (TRK) (or oscillator strength) sum rule \cite{Bethe}. While the TRK sum rule is discrete and derived from quantum-mechanical arguments, it is equivalent to the causality-based integral $f-$sum rule, which relates the integral of the imaginary part of the linear susceptibility, over the entire electromagnetic spectrum, to the total electron density \cite{Ziman,Lucarini}. One can derive the TRK sum rule from the $f-$sum rule by discretizing the integral through a physically motivated oscillator representation of the linear susceptibility, as further discussed in Section (\ref{sec:basis}) and in Ref. \cite{shim2021fundamental}, where the discretized causality-based sum rule was used to bound the linear optical response of any transparent medium. The equivalence between these sum rules, derived from very different arguments, is nontrivial and surprising, with important implications for the analysis and design of new materials and metamaterials with extreme response.

In this work, considering the case of SHG, we explore a new connection between causality-based integral sum rules and quantum-mechanics-based discrete sum rules, unveiling new constraints on the transition dipole moments and various implications for the SHG process. This is achieved by considering the Kramers--Kronig relations enforced by the principle of causality for the SHG susceptibility, $\chi^{(2)}(\omega,\omega)$, in combination with a new representation basis for the imaginary part of the susceptibility deduced from the perturbation solutions to the Schr\"odinger equation. The validity of the new set of discrete sum rules derived here is also confirmed by an independent quantum mechanical analysis. As in the case of TRK and $f-$sum rules, two very different approaches lead to identical results. 

In the following, the Kramers--Kronig relations and associated integral sum rules for the SHG susceptibility are briefly reviewed in Section (\ref{sec:KK-sr}). In Section (\ref{sec:basis}), a new representation basis for the imaginary part of the SHG susceptibility is established and used to derive new discrete sum rules on the transition dipole moments. In Section (\ref{sec:quantum}), these sum rules are related to an independent quantum mechanical analysis, further proving their validity. In Section (\ref{sec:examples}), examples of two--level and three--level systems are given, discussing various implications of the derived sum rules for the SHG process in these systems. Finally, in Section (\ref{sec:discussion}) we summarize our results, discuss the significance of our findings, and highlight a few open questions.

\section{\label{sec:KK-sr}Kramers--Kronig relations and integral sum rules for ${\mathop{\rm Im}\nolimits} \left\{ {{\chi ^{(2)}}(\omega ,\omega )} \right\}$}

Scandolo and Bassani \cite{Scandolo1995} presented an elegant proof showing that ${\chi ^{(2)}}(\omega ,\omega )$, ${\omega ^2}{\chi ^{(2)}}(\omega ,\omega )$, and ${\omega ^4}{\chi ^{(2)}}(\omega ,\omega )$ are holomorphic in the upper half of the complex frequency plane and exhibit an asymptotic high-frequency decay faster than $\omega^{-1}$. As a result, several Kramers--Kronig relations for the SHG susceptibility can be derived (for brevity, only the relations for the real part of ${\chi ^{(2)}}(\omega ,\omega )$ are shown here):

\begin{equation}
 {\omega ^{2m}}{\mathop{\rm Re}\nolimits} \left\{ {{\chi ^{(2)}}(\omega ,\omega )} \right\} = \frac{2}{\pi }\dashint_0^\infty  {\frac{{{{\omega '}^{2m + 1}}{\mathop{\rm Im}\nolimits} \left\{ {{\chi ^{(2)}}(\omega ',\omega ')} \right\}}}{{{{\omega '}^2} - {\omega ^2}}}d\omega '},
 \label{KKSHG}
\end{equation}


\noindent where $m=0,1,2$ and $\dashint$ indicates the principal value integral. In the same work \cite{Scandolo1995}, the asymptotic behavior of the SHG susceptibility was also used to determine the following integral sum rules:

\begin{eqnarray}
\dashint_0^\infty  {{\omega }{\mathop{\rm Im}\nolimits} \left\{ {{\chi ^{(2)}}(\omega ,\omega )} \right\}d\omega }  = 0 , 
 \label{SRSHG1}
 \\
\dashint_0^\infty  {{\omega ^{3}}{\mathop{\rm Im}\nolimits} \left\{ {{\chi ^{(2)}}(\omega ,\omega )} \right\}d\omega }  =  0, 
 \label{SRSHG2}
\\
\dashint_0^\infty  {{\omega ^{5}}{\mathop{\rm Im}\nolimits} \left\{ {{\chi ^{(2)}}(\omega ,\omega )} \right\}d\omega }  =  -  {\frac{\pi }{{16}}\frac{{{e^3}N}}{{{m^3}}}\left\langle {\frac{{{\partial ^3}V}}{{\partial {x}^3}}} \right\rangle _{0} , } 
 \label{SRSHG3}
\end{eqnarray}

\noindent where $N$ represents the electron density, $e$ is the electron charge, $m$ is the electron mass, $V(x)$ denotes the potential experienced by the electrons, and the averaging of the third derivative in Eq. (\ref{SRSHG3}) is conducted over the ground state of the system. Interestingly, it is clear that the imaginary part of $\chi ^{(2)}(\omega,\omega)$ has to change sign along the electromagnetic spectrum in order to satisfy Eqs. (\ref{SRSHG1})--(\ref{SRSHG3}). In contrast to the imaginary part of  the linear susceptibility, however, ${\mathop{\rm Im}\nolimits} \left\{ {{\chi ^{(2)}}(\omega ,\omega )} \right\}$ does not directly correlate with optical absorption or gain. Instead, it denotes a phase relationship between nonlinear polarization and applied fields, without necessarily implying any time-averaged absorbed or gained power \cite{Bloembergen1965}.

In addition to Eqs. (\ref{KKSHG})--(\ref{SRSHG3}), Kramers--Kronig relations that calculate ${\mathop{\rm Im}\nolimits} \left\{ {{\chi ^{(2)}}(\omega ,\omega )} \right\}$ from ${\mathop{\rm Re}\nolimits} \left\{ {{\chi ^{(2)}}(\omega ,\omega )} \right\}$ can also be obtained, and an extra set of sum rules for ${\mathop{\rm Re}\nolimits} \left\{ {{\chi ^{(2)}}(\omega ,\omega )} \right\}$ was also established in \cite{Scandolo1995}. These theoretical results have been used to assist in the analysis of experimental data, facilitating the connection between the phase and amplitude of the susceptibilities \cite{Kishida1993,Lucarini2003}, and providing insight into the possible presence of additional contributions outside the measured frequency range. 

\section{\label{sec:basis}Basis for ${\mathop{\rm Im}\nolimits} \left\{ {{\chi ^{(2)}}(\omega ,\omega )} \right\}$ and new discrete sum rules}

A perturbative solution to the Schr\"ondiger equation gives the well-established sum-over-states expression for the nonlinear SHG susceptibility \cite{Orr1971,Stegeman}

\begin{equation}
\begin{array}{l}
{\chi ^{(2)}}(\omega ,\omega ) = \frac{{N{e^3}}}{{{\hbar ^2}}}\sum\limits_{n,m}{} ^{'} {\left[ {\frac{{{x_{0n}}{{\bar x}_{nm}}{x_{m0}}}}{{({\omega _{n0}} - 2\omega  - i{\gamma _n})({\omega _{m0}} - \omega  - i{\gamma _m})}} + } \right.} \\
\left. {\frac{{{x_{0n}}{{\bar x}_{nm}}{x_{m0}}}}{{({\omega _{n0}} + \omega  + i{\gamma _n})({\omega _{m0}} + 2\omega  + i{\gamma _m})}} + \frac{{{x_{0n}}{{\bar x}_{nm}}{x_{m0}}}}{{({\omega _{n0}} + \omega  + i{\gamma _n})({\omega _{m0}} - \omega  - i{\gamma _m})}}} \right],
\end{array}
\label{perturbation}
\end{equation}

\noindent where $x$ represents the position operator, ${x_{nm}} = \left\langle n \middle| x \middle| m \right\rangle$ denotes its $(n,m)$ element, $\bar x = x - x_{00}\mathbb{I}$ (i.e., $\bar x$ is the same as the position operator for its off-diagonal elements and the difference of the position operator with $x_{00}$ for its diagonal elements), and the notation $\sum\limits_{}{} ^{'}$ indicates that the ground state is excluded from the summation over the states, i.e., the summation is conducted solely over the excited states. The energy level differences are given by $E_{n0}=\hbar \omega_{n0}=\hbar(\omega_{n}-\omega_{0})$. As usually done, since the dipole moment of the molecule is proportional to the position ($\mu=-ex$), we refer to $x_{nm}$ as the transition dipole moments and to $x_{nn}$ as the dipole moment of the excited state. $\gamma_{n}$ is a linewidth parameter (decay rate) that phenomenologically models various broadening/damping mechanisms \cite{Boyd2020}. We assume that the system is initially in its ground state, denoted as $\left| 0 \right\rangle$.
 
Following \cite{Scandolo1995a}, using fractional decomposition, one can then rewrite Eq. (\ref{perturbation}) in a more compact, yet still general, form: 

\begin{equation}
\begin{aligned}
{\chi ^{(2)}}(\omega ,\omega ) = 
& \sum\limits_n{} ^{'} {\left[ {\frac{{\alpha _1^{(n)}}}{{({\omega _{n0}} - \omega  - i{\gamma _n})}} + \frac{{\alpha _2^{(n)}}}{{({\omega _{n0}} - 2\omega  - i{\gamma _n})}}} \right]} \\
& + \sum\limits_{n'}{} ^{'} {\frac{{\alpha _3^{(n')}}}{{{{({\omega _{n'0}} - \omega  - i{\gamma _{n'}})}^2}}} + {{\left( {\omega  \to  - \omega } \right)}^ * }}. 
\end{aligned}
\label{fraction}
\end{equation}
Here, $n'$ denotes doubly resonant conditions where either $\omega_{mn'}=\omega_{n'0}$ or $\omega_{m0}=\omega_{0n'}$ (note that, if ``$0$'' denotes the ground state, the latter resonant condition would correspond to a negative transition frequency, and hence to an ``antiresonance'' that can only be induced in an active system; however, our analysis will be limited to passive systems). $\left( {\omega  \to  - \omega } \right)^ *$ denotes the sum of the negative-frequency and complex-conjugate functions as calculated from the first three terms of (\ref{fraction}). The coefficients $\alpha_{1}^{(n)}$ and $\alpha_{2}^{(n)}$ are the nonlinear oscillator strengths associated with single- and two-photon resonances at $\omega_{n0}$, respectively, and $\alpha_{3}^{(n')}$ is the oscillator strength associated with the specific condition of a double-resonance. These nonlinear oscillator strengths are given by 

\begin{equation}
\begin{aligned}
\alpha _1^{(n)} = 
& \frac{{N{e^3}}}{{{\hbar ^2}}}\sum\limits_m {} ^{'} {\frac{{{\mathds{1}_{{\omega _{mn}} \ne {\omega _{n0}}}}}}{{({\omega _{m0}} - 2{\omega _{n0}})}}}  {x_{0m} \bar x_{mn} x_{n0}} \\ 
& + \frac{{N{e^3}}}{{{\hbar ^2}}}\sum\limits_m {} ^{'} {\frac{{{\mathds{1}_{{\omega _{m0}} \ne  - {\omega _{n0}}}}}}{{({\omega _{m0}} + {\omega _{n0}})}}}  {x_{0m} \bar x_{mn} x_{n0} }, 
\end{aligned}
\label{alpha_1}
\end{equation}

\begin{equation}
\alpha _2^{(n)} = \frac{{2N{e^3}}}{{{\hbar ^2}}}\sum\limits_m {} ^{'} {\frac{{{\mathds{1}_{{\omega _{nm}} \ne {\omega _{m0}}}}}}{{(2{\omega _{m0}} - {\omega _{n0}})}}}  {x_{0n} \bar x_{nm} x_{m0}  } ,
\label{alpha_2}
\end{equation}

\begin{equation}
\begin{aligned}
\alpha _3^{(n')} = 
& \frac{{N{e^3}}}{{2{\hbar ^2}}}{\mathds{1}_{{\omega _{mn'}} = {\omega _{n'0}}}} {x_{0m} \bar x_{mn'} x_{n'0} }  \\
& - \frac{{N{e^3}}}{{{\hbar ^2}}}{\mathds{1}_{{\omega _{m0}} =  - {\omega _{n'0}}}} {x_{0m} \bar x_{mn'} x_{n'0} } .
\end{aligned}
\label{alpha_3}
\end{equation}

The algebraic manipulation giving Eq. (\ref{fraction}) from Eq. (\ref{perturbation}) was initially introduced in \cite{Scandolo1995a}, albeit with the goal to study a single doubly resonant system (namely, with only one term for each of the summations in (\ref{fraction})). Instead, here we are interested in how the general form of Eq. (\ref{fraction}) suggests a possible basis to discretize the integral sum rules discussed in Section (\ref{sec:KK-sr}). To this end, in the limit of vanishing linewidth, by inspecting Eq. (\ref{fraction}) one can deduce the following expression for the imaginary part of the nonlinear susceptibility,

\begin{widetext}
\begin{equation}
\begin{aligned}
{\mathop{\rm Im}\nolimits} \{ {\chi ^{(2)}}(\omega ,\omega )\}  = 
& \sum\limits_n{} ^{'} {\mathop {\lim }\limits_{{\gamma _n} \to {0^ + }} \left( {{\mathop{\rm Re}\nolimits} \{ \alpha _1^{(n)}\} {f_{{\gamma _n}}}(\omega  - {\omega _{n0}}) + \frac{{{\mathop{\rm Re}\nolimits} \{ \alpha _2^{(n)}\} }}{2}{f_{{{{\gamma _n}} \mathord{\left/
 {\vphantom {{{\gamma _n}} 2}} \right.
 \kern-\nulldelimiterspace} 2}}}(\omega  - {{{\omega _{n0}}} \mathord{\left/
 {\vphantom {{{\omega _{n0}}} 2}} \right.
 \kern-\nulldelimiterspace} 2})} \right)}  \\ 
 & + \sum\limits_{n'}{} ^{'} {\mathop {\lim }\limits_{{\gamma _{n'}} \to {0^ + }} \left( {{\mathop{\rm Re}\nolimits} \{ \alpha _3^{(n')}\} {f'_{{\gamma _{n'}}}}(\omega  - {\omega _{n'0}})} \right)} ,
\label{basis_lorentzian}
\end{aligned}
\end{equation}
\end{widetext}

\noindent where $f_{\gamma_{n}}(\omega)$ is the Lorentzian distribution: ${f_{\gamma_{n}} }(\omega ) = \frac{\gamma_{n} }{{{\omega ^2} + {\gamma_{n} ^{2}}}}$, and $f_{\gamma}'$ is its frequency derivative. Note that the $1/2$ weight and the modified broadening $\gamma_{n}/2$ of the Lorentzian associated with two-photon resonances is in full agreement with the permutation symmetry of the nonlinear susceptibility \cite{Boyd2020}. Additionally, it is easily found that ${\lim _{\gamma_{n}  \to {0^ + }}}{f_{\gamma_{n} ,{\gamma_{n}  \mathord{\left/ {\vphantom {\gamma_{n}  2}} \right. \kern-\nulldelimiterspace} 2}}}(\omega ) = \pi \delta (\omega )$, (see, e.g., \cite{Balakrishnan}) and ${\lim _{\gamma_{n}  \to {0^ + }}}{f'_{\gamma_{n}} }(\omega ) = \pi \delta '(\omega )$, where $\delta $ is the Dirac--$\delta$ distribution and $\delta' $ its frequency derivative.  
One can then verify that substituting (\ref{basis_lorentzian}) into the Kramers-Kronig relation (\ref{KKSHG}) yields (\ref{fraction}) for the case of $\gamma_{n} = {0 }$. We also note that, in writing Eq. (\ref{basis_lorentzian}), we made the assumption that only the real part of the coefficients $\alpha_{1}^{(n)}$, $\alpha_{2}^{(n)}$ and $\alpha_{3}^{(n')}$ enter this expression 
, essentially neglecting the effect of the transition dipole moments' phases on the imaginary part of the susceptibility. Rather surprisingly, while this assumption is strictly valid only in the off-resonance case, $\omega  \ll {\omega _{n0}}$ (which is important, per se, in many ultrafast applications), it ultimately leads to sum rules that are completely general, as proven by an independent quantum mechanical analysis in Section (\ref{sec:quantum}).

The considered expression for the imaginary part of the SHG susceptibility, in the limit of vanishing linewidth, involves delta functions and their frequency derivative. Specifically, there are two delta functions for each resonance frequency, $\omega_{n0}$, corresponding to single-photon and two-photon resonances, as well as the possible inclusion of the derivative of a delta function for the doubly resonant case. This is intriguingly different from the linear case, where the imaginary part of the linear susceptibility, in the limit of zero loss, can be expressed as a sum of only delta functions \cite{Ziman}. In Ref. \cite{shim2021fundamental}, this fact was used to derive a general representation of the linear susceptibility by discretizing the Kramers-Kronig relation for the real part of $\chi ^{(1)}$ using delta functions as localized basis functions for the imaginary part of $\chi ^{(1)}$. Since the coefficients of this representation are proportional to the oscillator strengths, it was then possible to derive an upper bound for ${\mathop{\rm Re}\nolimits} \left\{ {{\chi ^{(1)}}} \right\}$  by using the $f-$sum rule. While the same mathematical trick, using a basis of delta functions, could be applied to discretize the nonlinear Kramers-Kronig relation in Eq. (\ref{KKSHG}) (any collocation methods with localized basis functions would work), this would not give much insight as the resulting coefficients of this representation would not be directly relatable to the nonlinear oscillator strengths and transition dipole moments, in contrast with the linear case. In fact, the general expression for the imaginary part of $\chi ^{(2)}(\omega ,\omega )$, given by Eq. (\ref{basis_lorentzian}), suggests that a more natural choice for localized basis functions for $\chi ^{(2)}(\omega ,\omega )$ should include both delta functions and their derivatives, such that the coefficients resulting from this discretization would now be directly related to the nonlinear oscillator strengths, as shown by Eqs. (\ref{alpha_1}),(\ref{alpha_2}),(\ref{alpha_3}). Most importantly, this approach can also yield new sum rules on the transition dipole moments. Specifically, inserting the basis implied by Eq. (\ref{basis_lorentzian}) into the causality-based integral sum rules in Eqs. (\ref{SRSHG1})--(\ref{SRSHG3}) yields a discretized version of these sum rules in terms of the nonlinear oscillator strengths,

\begin{widetext}

\begin{equation}
\sum\limits_n{} ^{'} {\left( {{\omega _{n0}}{\mathop{\rm Re}\nolimits}\{\alpha _1^{(n)}\} + \frac{{{\omega _{n0}}}}{4}{\mathop{\rm Re}\nolimits}\{\alpha _2^{(n)}\}} \right) - \sum\limits_{n'}{} ^{'} {\mathop{\rm Re}\nolimits}\{{\alpha _3^{(n')}}\} }  = 0,
\label{strengths_1}
\end{equation}

\begin{equation}
\sum\limits_n{} ^{'} {\left( {\omega _{n0}^3{\mathop{\rm Re}\nolimits}\{\alpha _1^{(n)}\} + \frac{{\omega _{n0}^3}}{{16}}{\mathop{\rm Re}\nolimits}\{\alpha _2^{(n)}\}} \right) - \sum\limits_{n'}{} ^{'} {3\omega _{n'0}^2{\mathop{\rm Re}\nolimits}\{\alpha _3^{(n')}\}} }  = 0,
\label{strengths_2}
\end{equation}

\begin{equation}
\sum\limits_n{} ^{'} {\left( {\omega _{n0}^5{\mathop{\rm Re}\nolimits}\{\alpha _1^{(n)}\} + \frac{{\omega _{n0}^5}}{{64}}{\mathop{\rm Re}\nolimits}\{\alpha _2^{(n)}\}} \right) - \sum\limits_{n'}{} ^{'} {5\omega _{n'0}^4{\mathop{\rm Re}\nolimits}\{\alpha _3^{(n')}\}} }  =  - \frac{{{e^3}N}}{{{16m^3}}}\left\langle {\frac{{{\partial ^3}V}}{{\partial {x}^3}}} \right\rangle _{0}.
\label{strengths_3}
\end{equation}

\end{widetext}

While passivity implies that the transition frequencies are positive, $\omega_{n0}>0$, for all energy levels, it does not constrain the real part of the nonlinear oscillator strengths ${\mathop{\rm Re}\nolimits} \{ \alpha \} $ to take only positive values, in contrast with the linear case where the oscillator strengths are always positive.  If we then combine Eqs. (\ref{alpha_1})--(\ref{alpha_3}) with (\ref{strengths_1})--(\ref{strengths_3}) and after some algebraic manipulations, we obtain a new set of discrete sum rules for the transition dipole moments:

\begin{widetext}

\begin{equation}
\sum\limits_{n,m}{} ^{'} { \frac{{{\omega _{m0}} - {\omega _{n0}}}}{{{\omega _{m0}} + {\omega _{n0}}}}{\mathop{\rm Re}\nolimits} \{ {x_{0m}}{\bar x_{mn}}{x_{n0}}\} }  = 0,
\label{sum1}
\end{equation}

\begin{equation}
\sum\limits_{n,m}{} ^{'} \frac{{\omega _{m0}^3 + 3\omega _{m0}^2{\omega _{n0}} + 6\omega _{n0}^2{\omega _{m0}} - 4\omega _{n0}^3}}{{{\omega _{m0}} + {\omega _{n0}}}}{\mathop{\rm Re}\nolimits} \{ {x_{0m}}{\bar x_{mn}}{x_{n0}}\}   = 0,
\label{sum2}
\end{equation}

\begin{equation}
\sum\limits_{n,m}{} ^{'} \frac{{\omega _{m0}^5 + 3\omega _{m0}^4{\omega _{n0}} + 6\omega _{m0}^3\omega _{n0}^2 + 12\omega _{m0}^2\omega _{n0}^3 + 24\omega _{n0}^4{\omega _{m0}} - 16\omega _{n0}^5}}{{{\omega _{m0}} + {\omega _{n0}}}}{\mathop{\rm Re}\nolimits} \{ {x_{0m}}{\bar x_{mn}}{x_{n0}}\} = \frac{{2{\hbar ^2}}}{{{m^3}}}\left\langle {\frac{{{\partial ^3}V}}{{\partial {x}^{3}}}} \right\rangle _{0} .
\label{sum3}
\end{equation}
\end{widetext}

Although (\ref{sum1}) is automatically fulfilled (all $n=m$ terms are zero, whereas for $n \ne m$ the pairs $(n,m)$ and $(m,n)$ cancel each other), the second and third sum rules (\ref{sum2}), (\ref{sum3}) provide a set of new constraints that the transition dipole moments must satisfy to be consistent, ultimately, with the principle of causality, provided that our discretization approach is justified. In the next section, we show that the new sum rules (\ref{sum2}), (\ref{sum3}) can, in fact, be derived from an independent quantum mechanical analysis, confirming the deep connection between causality-based and quantum-mechanics-based constraints also in the nonlinear optical case, and proving that our proposed basis to represent the imaginary part of the susceptibility is a physically sensible choice to discretize causality-based integral relations.

\section{\label{sec:quantum}Independent Quantum Mechanical Derivation}

In this Section, we connect the derived sum rules, as established by the principle of causality and the sum-over-states expression of the SHG susceptibility, with an independent quantum-mechanical analysis. To do this, we employ generalized quantum mechanical sum rules, based on the operator theory of quantum mechanics.

\subsection{Quantum Mechanical Analysis of Sum Rule (\ref{sum2})}

Sum rule (\ref{sum2}) can be alternatively obtained from the generalized Thomas--Reiche--Kuhn (TRK) sum rules, which were derived for the calculation of the fundamental limits of the off-resonance nonlinear optical response \cite{Kuzyk2000,Kuzyk2000a,Kuzyk2001} and were additionally used for predicting the frequency dispersion of the hyperpolarizabilities spectrum \cite{Hu2010} and, very recently, for the derivation of fundamental bounds on resonant nonlinear optical responses in multi-quantum-well systems \cite{Hao2024}. These generalized TRK sum rules can be written as

\begin{equation} \label{TRK}
\sum_n  {\left[ {{\omega _n} - {\textstyle{1 \over 2}}({\omega _p} + {\omega _q})} \right]x_{pn} x_{nq} = {\textstyle{{Z\hbar } \over {2m}}}{\delta _{pq}}},
\end{equation}

\noindent where $\delta_{pq}$ is the Kronecker delta function and $Z$ is the number of electrons in the system. We take $q=0$ and consider all possible values of $p \in {\mathbb{Z}^ + }\backslash \{ 0\} $  in (\ref{TRK}). By multiplying both sides by $\omega_{n0}x_{n0}$ and summing over all $n \in {\mathbb{Z}^ + }\backslash \{ 0\} $, we obtain

\begin{align} \label{2nd sum rule 1}
\begin{split}
&\sum_n{}^{'} \omega_{n 0}^2 x_{0 n} \bar{x}_{n n} x_{n 0} + \\ 
&  \sum_{n,p\neq n}{}^{'} (4\omega_{n0}\omega_{p0}-\omega_{n0}^2-\omega_{p0}^2)x_{0n}x_{np}x_{p0}=0.
\end{split}
\end{align}

\noindent Summing (\ref{2nd sum rule 1}) with its index-interchanged expression, and taking into account that $x_{ij}=x_{ji}^*$, gives

\begin{align} \label{2nd sum rule 2}
\begin{split}
&\sum_n{}^{'} \omega_{n 0}^2 x_{0 n} \bar{x}_{n n} x_{n 0} + \\ 
&  \sum_{n,p\neq n}{}^{'} (4\omega_{n0}\omega_{p0}-\omega_{n0}^2-\omega_{p0}^2){\mathop{\rm Re}\nolimits}(x_{0n}x_{np}x_{p0})=0.
\end{split}
\end{align}

\noindent It is then straightforward to see that (\ref{2nd sum rule 2}) is just a rearrangement of (\ref{sum2}).

\subsection{Quantum Mechanical Analysis of Sum Rule (\ref{sum3})}

The independent derivation of sum rule (\ref{sum3}) is more involved. First, we consider the following commutation relations between momentum and Hamiltonian: $[p, H]=-i \hbar \frac{\partial V}{\partial x}$, $[p,[p, H]]=(-i \hbar)^2 \frac{\partial^2 V}{\partial x^2}$ and $[p,[p,[p, H]]]=(-i \hbar)^3 \frac{\partial^3 V}{\partial x^3}$, where $V(x)$ is the potential function of the system. Additionally, $[p,[p,[p, H]]]=-3 p[p, H] p$.

Similar to \cite{Wang1999}, we consider a generalized sum rule:

\begin{align}
\begin{split}
& \sum_{l, m}\left(E_l-E_m\right)\langle 0|p| l\rangle\langle l| p |m\rangle\langle m|p| 0\rangle \\
& =\sum_{l, m}\langle 0|p| l\rangle\langle l|[H, p]| m\rangle\langle m|p| 0\rangle \\
& =\langle 0|p[H, p] p| 0\rangle.
\end{split}
\end{align}

\noindent Then, since $[x, H]=\left[x, \frac{p^2}{2 m}\right]=\frac{i \hbar}{m} p$, we have: $\langle n|[x, H]| m\rangle=x_{n m}E_{mn}=\frac{i \hbar}{m} p_{n m}$. Hence, we get $\left\langle 0 \right|[p,[p,[p,H]]]\left| 0 \right\rangle  = {( - i\hbar )^3}{\left\langle {\frac{{{\partial ^3}V}}{{\partial {x^3}}}} \right\rangle _0}$ and

\begin{align} \label{3rd sum rule}
\begin{split}
&\frac{{{\hbar ^2}}}{{{m^3}}}{\left\langle {\frac{{{\partial ^3}V}}{{\partial {x^3}}}} \right\rangle _0} = 3\sum\limits_{l,m} {{{{\omega _{lm}^{2}} }}{\omega _{l0}} {\omega _{m0}} {\mathop{\rm Re}\nolimits} \left( {{x_{0l}}{x_{lm}}{x_{m0}}} \right)} .
\end{split}
\end{align}

Our objective here is to demonstrate the equivalence between (\ref{sum3}) and (\ref{3rd sum rule}). To achieve this, we perform the following algebraic manipulations to (\ref{sum3}). The sum rule (\ref{sum3}) can be expressed as:

\begin{equation} 
\frac{{2{\hbar ^2}}}{{{m^3}}}{\left\langle {\frac{{{\partial ^3}V}}{{\partial {x^3}}}} \right\rangle _0} = \frac{1}{2} (\sum_{n,m\neq n}{}^{'}  A_{mn} + \sum_{n}{}^{'} B_{nn} + \sum_{m}{}^{'} B_{mm}),
\label{SR1}
\end{equation}

\noindent where ${A_{nm}} = (-15\omega _{m0}^4 -15\omega _{n0}^4 + 42\omega _{m0}^3{\omega _{n0}} +42\omega _{n0}^3{\omega _{m0}} - 24\omega _{m0}^2\omega _{n0}^2){\mathop{\rm Re}\nolimits}({x_{0m}}\bar {x}_{mn}{x_{n0}})$  and $B_{nn} = 15\omega _{n0}^4 x_{0n} \bar x_{nn} x_{n0}$.

Again using similar algebraic manipulations as before, we take the TRK sum rules (\ref{TRK}), for $q=0$ and all possible values of $p \in {\mathbb{Z}^ + }\backslash \{ 0\} $ and multiply both sides by $\omega^{3}_{n0}x_{n0}$. We denote $C_{pn}=-15\omega_{n0}^3(\omega_{p0}+\omega_{pn}){\mathop{\rm Re}\nolimits}(x_{0p}x_{pn}x_{n0})$, so that $B_{nn}=\sum_{p\neq n}^{} C_{pn}$. Then (\ref{SR1}) becomes:

\begin{equation} 
\frac{{2{\hbar ^2}}}{{{m^3}}}{\left\langle {\frac{{{\partial ^3}V}}{{\partial {x^3}}}} \right\rangle _0} = \frac{1}{2}\sum\limits_{n,m \ne n}{}^{'} {\left( {{A_{mn}} + {C_{mn}} + {C_{nm}}} \right)},
\label{SR2}
\end{equation}

\noindent where each of the summed terms of (\ref{SR2}) is simplified to $A_{mn}+C_{mn}+C_{nm}=12\omega_{n0}\omega_{m0}\omega_{nm}^{2} {\mathop{\rm Re}\nolimits}(x_{0n}x_{nm}x_{m0})$. The equivalence of (\ref{3rd sum rule}) with (\ref{SR2}) is then clear.

Rather strikingly, these results show that, just as in the linear case for the standard TRK sum rule, these new sum rules can be derived independently from either a quantum mechanical analysis or from causality considerations combined with a suitable discretization of the relevant integrals to connect the discretization coefficients to the transition dipole moments (\ref{basis_lorentzian}). We stress that, although the integral sum rules have been derived from the causality properties of the SHG susceptibility, their discrete form in terms of transition dipole moments, which rely on generic quantum mechanical operator theory [see Eqs. (\ref{TRK}) and (\ref{3rd sum rule})], is general and not limited to any specific nonlinear process. Some intriguing consequences of these sum rules are discussed next.

\section{\label{sec:examples}Examples}

\subsection{Two--level approximation}

It is often convenient to approximate the nonlinear response of a material using a two-level system \cite{Boyd2020}. In this case, the system can be described in its simplest form as having only two levels: a ground state and an excited state. Despite its simplicity, this model has proven valuable for understanding susceptibility trends in many-level systems \cite{Stegeman}. In a many-level system, the transition dipole moment to the first excited state can reach its maximum value when transitions to all states beyond the first excited state are zero, per Eq. (\ref{TRK}) (when $p$ and $q$ are both zero, the sum rule yields: ${\left| {{x_{10}}} \right|^2} = \frac{\hbar }{{2m{\omega _{10}}}}Z - \sum\nolimits_{n = 2}^\infty  {\frac{{{\omega _{n0}}}}{{{\omega _{10}}}}} {\left| {{x_{n0}}} \right|^2}$, which is maximized when the second term on the right-hand-side is zero, i.e., when there are only two levels). However, this does not necessarily imply that a two-level system exhibits the strongest possible nonlinear response compared to higher-level systems, as the nonlinear oscillator strengths are not simply proportional to $\left| {{x_{n0}}} \right|^2$ [see Eqs. (\ref{alpha_1})--(\ref{alpha_3})] and are not constrained by sign restrictions. 

Under the two--level approximation and assuming negligible linewidth/damping, the SHG susceptibility is given by \cite{Stegeman}

\begin{equation}
\label{two-levels}
\chi ^{(2)}(\omega ,\omega ) = \frac{{3N{e^3}}}{{{}{\hbar ^2}}}\frac{{\omega _{10}^2\bar x_{11}{\left| {x_{10}} \right|^2}}}{{(\omega _{10}^2 - {\omega ^2})(\omega _{10}^2 - 4{\omega ^2})}}.
\end{equation}

\noindent Interestingly, using the sum rule given by Eq. (\ref{sum3}), we can relate the transition dipole moments to the expected value of the third derivative of the potential function: ${\bar x_{11}}{\left| {{x_{10}}} \right|^2} = \frac{{2\hbar^2 }}{{15{m^3}\omega _{10}^4}}{\left\langle {\frac{{{\partial ^3}V}}{{\partial {x^3}}}} \right\rangle _0}$. This allows us to determine a new expression for the SHG susceptibility within the two-level approximation:

\begin{equation}
\chi ^{(2)}(\omega ,\omega ) = \frac{{2N{e^3}}}{{5{}{m^3}\omega _{10}^2}}\left\langle {\frac{{{\partial ^3}V}}{{\partial {x^3}}}} \right\rangle _{0} \frac{1}{{(\omega _{10}^2 - {\omega ^2})(\omega _{10}^2 - 4{\omega ^2})}}.
\label{2-level}
\end{equation}

This model suggests an interesting scaling law with respect to the transition frequency that is not immediately evident in an expression like Eq. (\ref{two-levels}) due to the dependence of the dipole moments on the transition frequencies. In the low-frequency limit, far from resonance, the susceptibility scales very rapidly with respect to the transition frequency, $\chi ^{(2)}(0 ,0 ) \propto \omega _{10}^{-6} $. We note however that, while the two-level approximation can be a useful simplification, such a model for the nonlinear susceptibility does not satisfy causality and, therefore, cannot be used to derive universal physical bounds and general scaling laws. Specifically, the sum rule (\ref{sum2}) is not satisfied, and the high-frequency asymptotic behavior follows $\sim{\omega ^{ - 4}}$, whereas the expected asymptotic behavior for the SHG susceptibility should be $\sim{\omega ^{ - 6}}$ \cite{Scandolo1995,Lucarini}.

\begin{figure}
\centering\includegraphics[width=0.95\linewidth, keepaspectratio]{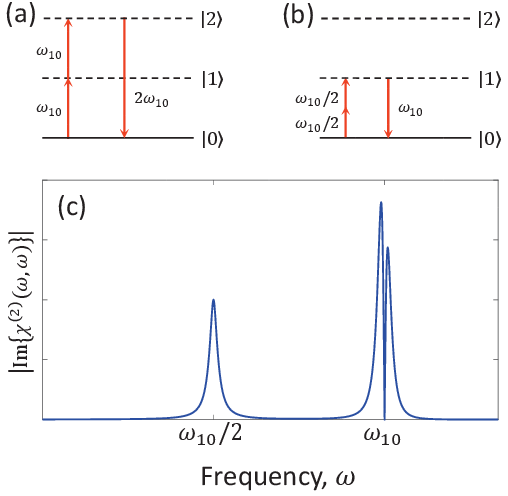}
\caption{\label{fig:fig_1} (a) Schematic representation of level transitions related to the double-resonance at $\omega=\omega_{10}$. (b) Schematic representation  of level transitions related to the two-photon resonance at $\omega=\omega_{10}/2$. (c) Absolute value of the imaginary part of the second-harmonic-generation susceptibility for a doubly resonant three--level system.}
\end{figure}

\subsection{Doubly resonant three--level system}

A three-level system is the simplest model for which the SHG susceptibility can fully satisfy the relevant Kramers-Kronig relations, and therefore causality.
In a scenario where the levels are equally spaced, resulting in resonance frequencies at $\omega_{10}$ and $\omega_{20}=2\omega_{10}$, the TRK sum rules (\ref{TRK}) give: $\bar{x}_{22}=0$ and $x_{01}\bar{x}_{11}=-3x_{02}x_{21}$. The nonlinear oscillator strength at $\omega_{20}$ is therefore zero, and the susceptibility doesn't present a resonance at this frequency. If the transition frequencies are positive (e.g., in a passive system), one can then identify two possible resonant conditions: one when the applied field frequency coincides with the first transition frequency $\omega = \omega _{10}$, leading to a double-resonance, as depicted in Fig.\ \ref{fig:fig_1}(a), and a second one when $\omega = \omega_{10}/2$, corresponding to a two-photon resonance, as depicted in Fig.\ \ref{fig:fig_1}(b). A plot of the absolute value of the imaginary part of the SHG susceptibility for a doubly resonant three--level system is depicted in Fig.\ \ref{fig:fig_1}(c). This consists of two Lorentzian functions at the transition frequency, $\omega_{10}$, and at half this frequency, $\omega_{10}/2$, as well as the derivative of a Lorentzian function at $\omega_{10}$, corresponding to the doubly resonance phenomenon, consistent with our discussion in Section \ref{sec:basis}. 

\begin{figure}
\centering\includegraphics[width=0.9\linewidth, keepaspectratio]{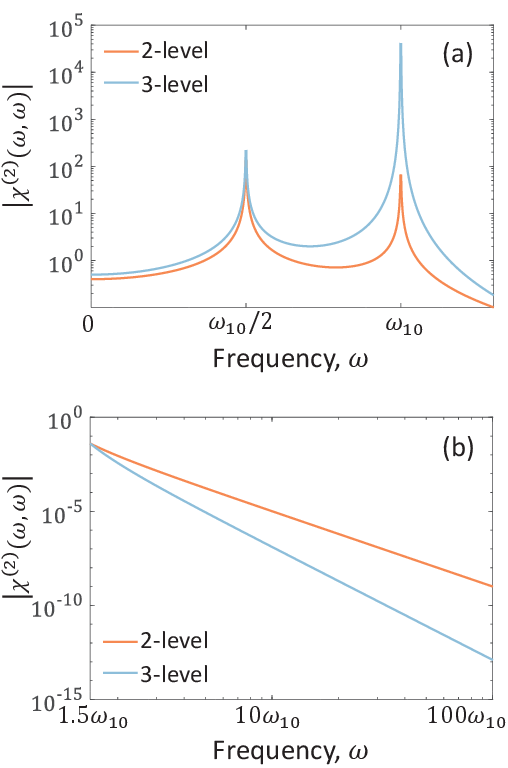}
\caption{\label{fig:fig_2} Absolute value of the SHG susceptibility in logarithmic scale for two--level and three--level systems calculated using Eqs. (\ref{2-level}) and (\ref{3-level}), in the (a) low-frequency and near-resonance regime and (b) high-frequency regime. Both formulas predict resonances at $\omega = \omega_{10}$ and $\omega = \omega_{10}/2$ (although the two--level system does not predict a double-resonance). The three--level system results in stronger nonlinear response near-resonance and in the low-frequency limit, while the two--level system results in stronger nonlinear response in the high-frequency limit (such result is, however, physically incorrect, as discussed in the main text).}
\end{figure}

For a doubly resonant three-level system with vanishing linewidth, the SHG susceptibility given by Eq. (\ref{fraction}) can then be simplified to:

\begin{equation}
\chi^{(2)}(\omega ,\omega ) = \frac{{\alpha _1}}{{{\omega _{10}} - \omega }}  + \frac{{\alpha _2}}{{{\omega _{10}} - 2\omega }}  + \frac{{\alpha _3}}{{{{({\omega _{10}} - \omega )}^2}}} + {(\omega  \to  - \omega )^ * },
\end{equation}

\noindent where $\alpha _1 =  - \frac{{N{e^3}}}{{2{}{\hbar ^2}}}\frac{{{{\left| {{x_{01}}} \right|}^2}{{\bar x}_{11}}}}{{{\omega _{10}}}} + \frac{{N{e^3}}}{{3{}{\hbar ^2}}}\frac{{{x_{01}}{x_{12}}{x_{20}}}}{{{\omega _{10}}}}$, $\alpha _2 = \frac{{2N{e^3}}}{{{}{\hbar ^2}}}\frac{{{{\left| {{x_{01}}} \right|}^2}{{\bar x}_{11}}}}{{{\omega _{10}}}} + \frac{{2N{e^3}}}{{3{}{\hbar ^2}}}\frac{{{x_{01}}{x_{12}}{x_{20}}}}{{{\omega _{10}}}}$ and  $\alpha _3 = \frac{{N{e^3}}}{{2{}{\hbar ^2}}}{x_{02}}{x_{21}}{x_{10}}$. 
One can then use the sum rule (\ref{sum3}) to obtain: ${\left| {{x_{01}}} \right|^2}{\bar x_{11}} =  - \frac{{{\hbar ^2}}}{{4{m^3}\omega _{10}^4}}\left\langle {\frac{{{\partial ^3}V}}{{\partial {x^3}}}} \right\rangle _0$ and ${x_{02}}{x_{21}}{x_{10}} = \frac{{{\hbar ^2}}}{{12{m^3}\omega _{10}^4}}\left\langle {\frac{{{\partial ^3}V}}{{\partial {x^3}}}} \right\rangle _0$. The susceptibility can then be written in terms of the linear susceptibility  for a single resonance ${\chi ^{(1)}}(\omega )$, as first shown in \cite{Scandolo1995a},

\begin{equation}
\begin{array}{l}
{\chi ^{(2)}}(\omega ,\omega ) =  - \frac{{N{e^3}}}{{2{m^3}}}{\left\langle {\frac{{{\partial ^3}V}}{{\partial {x^3}}}} \right\rangle _0}\frac{1}{{{{(\omega _{10}^2 - {\omega ^2})}^2}(\omega _{10}^2 - 4{\omega ^2})}}\\
  =  - \frac{{{{\left\langle {\frac{{{\partial ^3}V}}{{\partial {x^3}}}} \right\rangle }_0}}}{{2{N^2}{e^3}}}{\chi ^{(1)}}(2\omega ){\chi ^{(1)}}(\omega ){\chi ^{(1)}}(\omega ),
\end{array}
\label{3-level}
\end{equation}

\noindent which provides an analytical expression for the empirical relation between nonlinear and linear susceptibilities known as Miller's rule \cite{Miller1964}, and a specific definition for the proportionality constant (Miller's constant). The derived Miller's constant is consistent with findings from the anharmonic oscillator model, where the expected value of the third derivative of the potential at the ground level is substituted by the third derivative of the potential at the equilibrium position of a simple oscillator model \cite{Bassani1998}. We stress, however, that Miller's rule, which can be derived rigorously for a three-level system as shown here, can only provide approximate and qualitative predictions for more complex molecules and solid-state materials, since realistic materials involve more than three levels. Nevertheless, Eq. (\ref{3-level}) is expected to be approximately valid far from the transition frequencies, for example in the low-frequency regime, where it predicts the same scaling law as the two-level model, $\chi ^{(2)}(0 ,0 ) \propto \omega _{10}^{-6} $. In particular, if we compare (\ref{2-level}) with (\ref{3-level}), the non-causal two-level system approximates well the predictions of the causal three-level system in the low-frequency limit $\omega  \to 0$: $\left| {{{\chi _{{\rm{2 - level}}}^{(2)}} \mathord{\left/
 {\vphantom {{\chi _{{\rm{2 - level}}}^{(2)}} {\chi _{{\rm{3 - level}}}^{(2)}}}} \right.
 \kern-\nulldelimiterspace} {\chi _{{\rm{3 - level}}}^{(2)}}}} \right| = {4 \mathord{\left/
 {\vphantom {4 5}} \right.
 \kern-\nulldelimiterspace} 5}$. Fig.\ \ref{fig:fig_2} shows the absolute value of the susceptibilities calculated using equations (\ref{2-level}) and (\ref{3-level}), both normalized in the same way. The three--level formula (\ref{3-level}) predicts a stronger nonlinear response almost everywhere, especially near the double resonance at $\omega = \omega_{10}$, where the two-level model only predicts a single resonance. In general, the strongest SHG nonlinearity occurs in this doubly resonant scenario, with a clear trade-off between strength of nonlinearity and dispersion/bandwidth, which is similar to the linear case \cite{shim2021fundamental}, but with this tradeoff explicitly dependent on the spatial derivatives of the potential function in addition to the electron density. Interestingly, Fig.\ \ref{fig:fig_2}(b) shows that the two--level formula (\ref{2-level}) exhibits a stronger nonlinear response in the high-frequency limit ($\omega  \to \infty $). However, this result is physically incorrect, as it violates the expected asymptotic behavior of the SHG susceptibility \cite{Lucarini} and the relevant Kramers-Kronig relations, as mentioned above.

One may also wonder whether the derived expressions, constrained by the derived sum rules, may allow establishing fundamental limits on the second-order susceptibilities of three-level systems, perhaps improving upon existing bounds \cite{Kuzyk2000,Kuzyk2000a,Kuzyk2001}, and accounting for the tradeoffs with dispersion and bandwidth mentioned above, as done in \cite{shim2021fundamental} for the linear susceptibility. However, as is clear from Eq. (\ref{3-level}), bounding the nonlinear susceptibility would reduce to the challenging task of determining limitations to the expected value of the third derivative of the potential at the ground state. Moreover, it is hard in general to account for passivity constraints due to the non-positivity of the nonlinear oscillator strengths in the considered representation. For example, in the case considered above, the nonlinear oscillator strengths can be explicitly found as $\alpha _1 = \frac{{11}}{{72}}\frac{{{e^3}N}}{{{}{m^3}\omega _{10}^5}}\left\langle {\frac{{{\partial ^3}V}}{{\partial {x^3}}}} \right\rangle_0 $, $\alpha _2 =  - \frac{4}{9}\frac{{{e^3}N}}{{{}{m^3}\omega _{10}^5}}\left\langle {\frac{{{\partial ^3}V}}{{\partial {x^3}}}} \right\rangle _0 $ and $\alpha _3 = \frac{1}{{24}}\frac{{{e^3}N}}{{{}{m^3}\omega _{10}^4}}\left\langle {\frac{{{\partial ^3}V}}{{\partial {x^3}}}} \right\rangle_0$. Clearly, these oscillator strengths do not have the same sign and the imaginary part of the SHG susceptibility may take both positive and negative values throughout the frequency spectrum, consistent with (\ref{SRSHG1}) and (\ref{SRSHG2}), without necessarily violating passivity.

We recently used a different approach to obtain fundamental limits for the second-order susceptibility of three-level systems, albeit without accounting for bandwidth/dispersion, using only the generalized TRK sum rules (\ref{TRK}), which automatically encode passivity restrictions, but no bandwidth information. Our results will be the subject of a future publication \cite{Hao2024}. Moreover, we note that the derivation of fundamental limits and scaling laws for higher-level systems will possibly require additional sum rules due to the increasing number of unknown nonlinear oscillator strengths. Specifically, such an analysis would necessitate sum rules that are orthogonal to the generalized TRK sum rules (\ref{TRK}) and the newly established SHG sum rule (\ref{sum3}) (whereas (\ref{sum2}) is dependent on the generalized TRK sum rules (\ref{TRK}) as we showed in Section \ref{sec:quantum}).


\section{\label{sec:discussion}Discussion}

In this paper, we have established a new ``theoretical bridge'' between causality-based integral sum rules for the second harmonic generation susceptibility and quantum-mechanics-based discrete sum rules that relate energy levels to transition dipole moments, the latter being valid for any linear and nonlinear optical materials. This connection is analogous to the well-established equivalence between the  integral $f-$sum rule for the linear susceptibility and the Thomas--Reiche--Kuhn sum rule \cite{Lucarini}, with our theoretical results highlighting how this equivalence between causality-based and quantum-mechanics-based sum rules encompass also nonlinear optical phenomena. 

The new set of sum rules presented here—derived both through a physically motivated discretization of the integral sum rules and from an independent quantum mechanical analysis—hold promise for validating measurements, theoretical models, and numerical simulations, as they provide strong constraints for the transition dipole moments and, hence, for the linear and nonlinear optical response of materials. Moreover, they may offer insights into the second-harmonic generation process across diverse scenarios, spanning from the characterization of molecular nonlinear susceptibilities and hyperpolarizabilities to investigating the optical responses of multi-quantum wells.

Looking ahead, we believe that an extension of our current analysis to third-order (and higher-order) harmonic generation susceptibilities is feasible, offering valuable insights into these important nonlinear optical processes and potentially unveiling new sum rules for the transition dipole moments. Moreover, an important question is whether the sum rules established in this work could be combined with other constraints to derive physically tight bounds on harmonic generation and, more generally, on the nonlinear optical response of systems with more than three levels (while in the linear case, the simplest causal system, namely, a single Lorentzian oscillator, saturating the total oscillator strength implied by the TRK sum rule, is optimal for maximizing the linear susceptibility \cite{shim2021fundamental}, there is no guarantee that a three-level system maximizes the second-order susceptibility in all scenarios).

Another interesting question is whether the developed basis for the imaginary part of the harmonic generation susceptibility, connecting the coefficients resulting from the discretization of causality-based integral relations to nonlinear oscillator strengths, could be used to develop a causality-based electromagnetic scattering theory similar to \cite{Zhang2023} but focused on nonlinear harmonic generation processes. This approach could establish a scattering representation that inherently incorporates causality, and therefore bandwidth and dispersion properties, potentially revealing significant constraints in the mathematical structure and physical behavior of harmonically generated scattered fields.

\begin{acknowledgments}

F.M. was supported by the Air Force Office of Scientific Research with Grant No. FA9550-22-1-0204. T.T.K. was supported by the Swiss National Science Foundation (SNSF), with Grant No.\ 203176, and the ``Stamatis G. Mantzavinos'' Postdoctoral Scholarship from the Bodossaki Foundation. H.L. and O.D.M. were supported by the Air Force Office of Scientific Research with Grant No. FA9550-22-1-0393, and by the Simons Collaboration on Extreme Wave Phenomena Based on Symmetries (Award No.\ SFI-MPS-EWP-00008530-09). F.M., H.L., and O.D.M. were also partially supported by the Air Force Research Laboratory, with Grant No. FA8650-16-D-5404.

\end{acknowledgments}



\bibliography{bibliography1}

\end{document}